\documentclass[trackchanges]{aastex7}
\usepackage{amsmath}
\usepackage{graphicx}
\usepackage{comment}
\usepackage{multirow}
\usepackage[caption=false]{subfig}

\begin{document}

\title{Predicting Associations between Solar Flares and Coronal Mass Ejections Using SDO/HMI Magnetograms and 
a Hybrid Neural Network
}


\author{Jialiang Li}
\affiliation{Institute for Space Weather Sciences, New Jersey Institute of Technology, University Heights, Newark, NJ 07102, USA}
\affiliation{Department of Computer Science, New Jersey Institute of Technology, University Heights, Newark, NJ 07102, USA}
\email{jl2356@njit.edu}

\author{Vasyl Yurchyshyn}
\affiliation{Big Bear Solar Observatory, New Jersey Institute of Technology, 40386 North Shore Lane, Big Bear City, CA 92314, USA}
\email{vasyl.yurchyshyn@njit.edu}

\author{Jason T. L. Wang}
\affiliation{Institute for Space Weather Sciences, New Jersey Institute of Technology, University Heights, Newark, NJ 07102, USA}
\affiliation{Department of Computer Science, New Jersey Institute of Technology, University Heights, Newark, NJ 07102, USA}
\email{wangj@njit.edu}

\author{Haimin Wang}
\affiliation{Institute for Space Weather Sciences, New Jersey Institute of Technology, University Heights, Newark, NJ 07102, USA}
\affiliation{Big Bear Solar Observatory, New Jersey Institute of Technology, 40386 North Shore Lane, Big Bear City, CA 92314, USA}
\affiliation{Center for Solar-Terrestrial Research, New Jersey Institute of Technology, University Heights, Newark, NJ 07102, USA}
\email{haimin.wang@njit.edu}

\author{Manolis K. Georgoulis}
\affiliation{Johns Hopkins University Applied Physics Laboratory, 
11100 Johns Hopkins Road, Laurel, MD 20723, USA}
\affiliation{Research Center for Astronomy and Applied Mathematics, Academy of Athens, 11527 Athens, Greece (on leave)}
\email{manolis.georgoulis@jhuapl.edu}

\author{Wen He}
\affiliation{Institute for Space Weather Sciences, New Jersey Institute of Technology, University Heights, Newark, NJ 07102, USA}
\affiliation{Center for Solar-Terrestrial Research, New Jersey Institute of Technology, University Heights, Newark, NJ 07102, USA}
\email{wen.he@njit.edu}

\author{Yasser Abduallah}
\affiliation{Institute for Space Weather Sciences, New Jersey Institute of Technology, University Heights, Newark, NJ 07102, USA}
\affiliation{Department of Computer Science, New Jersey Institute of Technology, University Heights, Newark, NJ 07102, USA}
\email{ya54@njit.edu}

\author{Hameedullah A. Farooki}
\affiliation{Department of Astrophysical Sciences,
Princeton University, Princeton, NJ 08544, USA} 
\email{hafarooki@princeton.edu}

\author{Yan Xu}
\affiliation{Institute for Space Weather Sciences, New Jersey Institute of Technology, University Heights, Newark, NJ 07102, USA}
\affiliation{Big Bear Solar Observatory, New Jersey Institute of Technology, 40386 North Shore Lane, Big Bear City, CA 92314, USA}
\affiliation{Center for Solar-Terrestrial Research, New Jersey Institute of Technology, University Heights, Newark, NJ 07102, USA}
\email{yan.xu@njit.edu}

\begin{abstract}
Solar eruptions, including flares and coronal mass ejections (CMEs), have a significant impact on Earth. 
Some flares are associated with
CMEs, and some flares are not.
The association between flares and CMEs is not always obvious.
In this study, we propose a new deep learning method, 
specifically a hybrid neural network (HNN) that combines
a vision transformer with long short-term memory,
to predict associations between flares and CMEs. 
HNN finds spatio-temporal patterns in the
time series of line-of-sight magnetograms of
solar active regions 
(ARs) collected by the Helioseismic and Magnetic Imager on board the Solar Dynamics Observatory and uses the patterns to predict whether a flare 
projected to occur
within the next 24 hours will be 
eruptive (i.e., CME-associated) or confined 
(i.e., not CME-associated).
Our experimental results demonstrate the good performance of 
the HNN method.
Furthermore, the results show that 
magnetic flux cancellation 
in polarity inversion line regions
may well play a role
in triggering flare-associated CMEs, a finding 
consistent with literature.
\end{abstract}

\section{Introduction} 
\label{sec:intro}

Solar flares are sudden and large explosions 
in the Sun. 
Some flares have a close relationship with
coronal mass ejections (CMEs), which
are observed in white-light coronagraph images \citep{2006ApJ...650L.143Y,2023ApJ...958L..34A}.
Considering the occurrence of CMEs, there are two types of flares \citep{2001ApJ...552..833M}.
One type is eruptive flares,
which are typically 
associated with
CMEs \citep{2007ApJ...665.1428W,2023ApJ...952..136A,2023ApJ...958..104K}.
The other type is non-eruptive or confined flares,
which are not associated with any CMEs
\citep{2015ApJ...801L..23T,2025ApJ...983..126T}.

CMEs can
release billions of tons of charged particles into space at high speeds
\citep{2000JGR...105.2375L,WH2012LRSP,2023ApJ...958L..34A}.
They have a profound impact on the near-Earth space environment and on Earth, with
potential consequences ranging from technological disturbances
to life-threatening scenarios \citep{2004SpWea...2.2004B}. 
As a result, significant efforts have been made to leverage new technologies for the early detection and forecasting of flares 
\citep{2015ApJ...798..135B,2017ApJ...843..104L,2018SoPh..293...28F,2018ApJ...856....7H,2018SoPh..293...48J,2018ApJ...858..113N,2019ApJ...877..121L,2020ApJ...891...10L,2020ApJ...895....3W,2022ApJ...941....1S,2022ApJS..263...28Z,2023NatSR..1313665A,2025ApJS..277...60W,2025ApJS..276...68X} and 
CMEs \citep{2008SoPh..248..471Q,2016ApJ...821..127B,2018ApJ...861..128I,2020ApJ...890...12L,2025ApJ...981...37Z}.
A recent review on the prediction of these solar energetic events can be found in 
\citet{Georgoulis2024}.
These new technologies are developed to better prepare and to mitigate the impacts of solar eruptions on Earth's technological systems.

Recently, machine learning 
\citep[and references therein]{alpaydin2016machine}
has drawn significant interest in the field.
This technology is able to learn from historical data to make predictions on new, unseen data. 
For instance, researchers employed machine learning algorithms and
vector magnetic data products from the
Helioseismic and Magnetic Imager \citep[HMI;][]{HMI} 
on board the Solar Dynamics Observatory \citep[SDO;][]{SDO}
to predict flares within the next 24 hours
\citep{2015ApJ...798..135B,2017ApJ...843..104L}. 
Others used similar machine learning algorithms and data to predict CMEs
\citep{2016ApJ...821..127B,2018ApJ...861..128I}.
More recently, deep learning \citep{LeCun2015}, a subfield of machine learning,
has emerged as a powerful tool to predict solar eruptions.
A suite of deep learning methods has been developed,
ranging from recurrent neural networks, including long short-term memory and gated recurrent units,
to convolutional neural networks \citep{1982PNAS...79.2554H,DBLP:journals/neco/HochreiterS97,LeCun2015}, to predict eruptive events
\citep{2019ApJ...877..121L,
2020ApJ...891...10L,
2020ApJ...890...12L,
2022ApJ...941....1S,
2023NatSR..1313665A,
2025ApJS..276...68X}. 

It is now widely believed that
flares and CMEs 
are manifestations of solar coronal relaxation in terms of excess magnetic energy or magnetic helicity  
\citep{1994PhPl....1.1684L,
1995A&A...304..585H,
1996ApJ...464L.199R,
2011LRSP....8....1C,
2012ApJ...753...88B}.
However, the association between them is not straightforward based on 
current observations \citep{yashiro_gopalswamy_2008,2018ApJ...869...99K}. 
Many efforts have been made to understand the role of the coronal magnetic field in the initialization, evolution, and eruption mechanisms in confined and eruptive flares
\citep{2005ApJ...630L..97T,2008ApJ...680..740D,2008ApJ...679L.151L,2018ApJ...853..105B}.
In order to improve the understanding of these physical mechanisms, research has also been conducted for CME predictions; e.g., some studies used physical parameters derived from photospheric vector magnetograms to
predict the occurrence of CMEs
\citep{2016ApJ...821..127B, 2018ApJ...861..128I,2020ApJ...890...12L,
2025ApJ...981...37Z}.

In this work, we attempt to further understand the
association between flares and CMEs
by harnessing the power of a vision transformer (ViT)
and long short-term memory (LSTM)
with SDO/HMI line-of-sight (LOS) magnetograms.
ViT \citep{DBLP:conf/iclr/DosovitskiyB0WZ21} is an effective deep learning method for image classification, while
LSTM \citep{2019ApJ...877..121L} works well for time series forecasting.
We develop a hybrid neural network, named HNN, that combines ViT and LSTM
to predict associations between flares and CMEs. 
HNN finds spatio-temporal patterns in the
time series of SDO/HMI LOS magnetograms of solar active regions (ARs), 
and uses the patterns to predict whether a flare 
projected to occur within the next 24 hours will be 
eruptive (i.e., CME-associated) or confined 
(i.e., not CME-associated).
The use of time series of HMI magnetogram images of ARs, 
processed through the well-trained HNN model, 
enables us to leverage the model's inherent strengths for pattern recognition and
spatio-temporal data analysis. 

The remainder of this paper is organized as follows.
Section \ref{sec:Data} describes the data used in our study.
Section \ref{sec:methodology} defines our prediction task
and presents the HNN model to tackle the task.
Section \ref{sec:Results} reports the experimental results. 
Section \ref{sec:conclusion} presents a discussion and concludes the paper.

\section{Data} \label{sec:Data}

Solar flares are classified according to 
their peak flux
(in watts per square meter, 
$\mbox{Wm}^{-2}$) of 1 to 8 \AA ~X-rays as
measured by the Geostationary Operational Environmental Satellite (GOES)
provided by the National Centers for Environmental
Information \citep{2019ApJ...877..121L,2023NatSR..1313665A}.
In this study, we consider 
three cumulative flare classes
\citep{2018SoPh..293...48J,2018ApJ...858..113N,2019ApJ...877..121L,2023NatSR..1313665A}:
$\geq$M5.0-class, $\geq$M-class and
$\geq$C-class. 
The $\geq$M5.0-class contains flares 
with peak soft X-ray flux above
$5 \times 10^{-5}\mbox{Wm}^{-2}$. 
Given its cumulative property, the $\geq$M-class contains GOES M- and X-class flares while the $\geq$C-class contains GOES C-, M- and X-class flares.
We select flares with
identified ARs that occurred in the period
between May 2010 and December 2022.
We then download line-of-sight (LOS) magnetograms
of Space-weather HMI
Active Region Patches \citep[SHARPs;][]{2014SoPh..289.3549B}, produced
by the SDO/HMI team, from the Joint Science
Operations Center.\footnote{\url{http://jsoc.stanford.edu/}}
These HMI magnetograms are downloaded at a frequency of 1 hour.
The downloaded magnetogram images have varying dimensions.
We crop these images to the same size with $448 \times 448$ pixels 
and then resize the images to
$224 \times 224$ pixels for model training purposes.
Following the literature
\citep{2021MNRAS.507.3519Z,2022ApJ...941....1S},
we exclude
ARs that are outside $\pm$$45^\circ$ of the central meridian to avoid projection effects.

Next, we extract data from the Space Weather
Database of Notifications, Knowledge, Information 
(DONKI),\footnote{\url{https://kauai.ccmc.gsfc.nasa.gov/DONKI/}}
to check whether a given flare 
is associated with a CME.
Combining information from DONKI, 
we create three datasets separately:
a $\geq$M5.0 dataset,
a $\geq$M dataset, and
a $\geq$C dataset.
In the $\geq$M5.0 dataset, 
there are 120 
$\geq$M5.0-class 
flares, 
among which 
69 flares 
are eruptive
and
51 flares are confined. 
In the $\geq$M dataset,
there are 619 
$\geq$M-class 
flares, 
comprising
184 eruptive flares and 435 confined flares.
In the $\geq$C dataset,
there are 773 
$\geq$C-class 
flares,
comprising
301 eruptive flares and 472 confined flares.
Note that, in practice, there are much more confined flares than eruptive flares in the $\geq$C dataset.
DONKI only records a 
fraction 
of those confined flares. 
Our data is mainly collected from DONKI.
Thus, we essentially use a undersampling method to tackle 
the imbalance issue in the $\geq$C dataset,
which helps avoid bias in model training.

\section{Methodology} \label{sec:methodology}

\subsection{Prediction Task} \label{subsec:prediction task}
We aim to tackle 
the following prediction task, which 
involves a binary classification problem.
Given a magnetogram $x_{t}$ at time point $t$ in an AR where the AR will produce a
$\gamma$-class flare within the next 24 hours of $t$,
we predict whether $x_{t}$ 
is positive or negative. 
Here, we consider $\gamma$ to be 
$\geq$M5.0, 
$\geq$M,
or $\geq$C, 
separately and individually.
Predicting $x_{t}$ to be positive means that 
the $\gamma$-class flare will be 
associated with
a CME (i.e., the flare is eruptive).
Predicting $x_{t}$ to be negative means that 
the $\gamma$-class flare will not be 
associated with
a CME (i.e., the flare is confined). 

\begin{figure}
    \centering
    \begin{minipage}{1\textwidth}
        \centering
        \includegraphics[width=\linewidth]{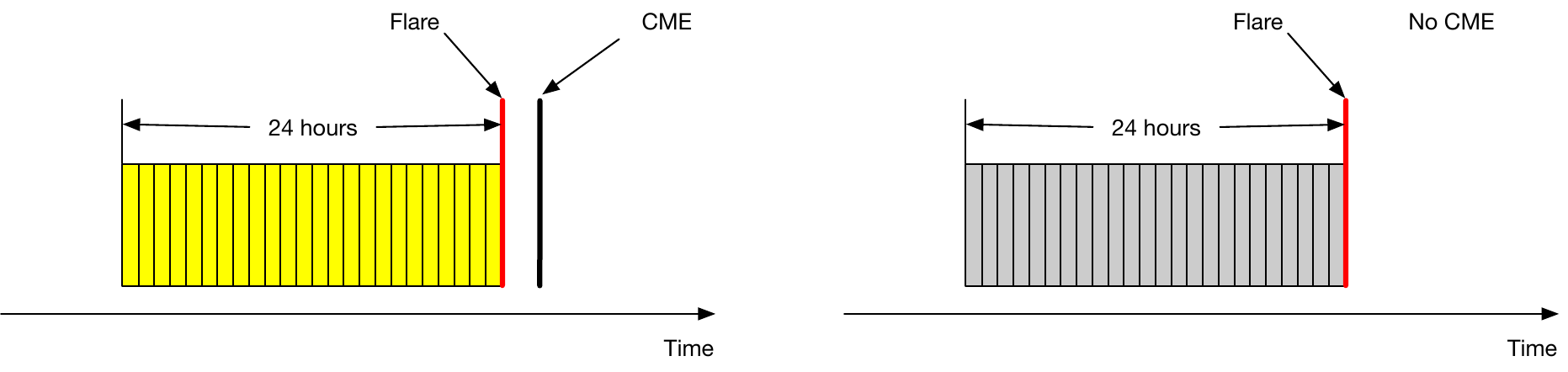} 
    \end{minipage}
    \caption{Construction of positive and negative magnetograms used in our prediction task. 
The magnetograms are collected at a frequency of  
1 hour.
Each rectangular box corresponds to 1 hour and contains 
one magnetogram.
The red vertical line indicates the peak time of a $\gamma$-class flare,
where $\gamma$ implies 
$\geq$M5.0, 
$\geq$M,
or $\geq$C.
The yellow rectangular boxes shown in the left panel 
contain magnetograms that are within the 24 hours prior to the peak time of 
a $\gamma$-class flare that is 
associated with
a CME; 
these yellow magnetograms belong to the positive class.
The gray rectangular boxes shown in the right panel
contain magnetograms that
are within the 24 hours prior to the peak time of a $\gamma$-class flare that is not 
associated with
a CME;
these gray magnetograms belong to the negative class.
Notice that in theory there are 
24 magnetograms that are within the 24 hours prior to the peak time of a $\gamma$-class flare.
However, in practice, since some magnetograms could be missing,
there could be fewer than 24 magnetograms 
collected in our datasets.
\label{fig:label_rule}}
\end{figure} 

Figure \ref{fig:label_rule} illustrates how we label the magnetograms at hand
for model training and testing.
We label the magnetograms for the 
$\geq$M5.0-, 
$\geq$M-,
and $\geq$C-class flares,
separately and individually.
For a flare that is 
associated with
a CME, 
we label the magnetograms collected 24 hours before the flare as positive,
as shown in the left panel of Figure \ref{fig:label_rule}.
For a flare that is not associated with
a CME, we label the magnetograms collected 24 hours before the flare as negative,
as shown in the right panel of Figure \ref{fig:label_rule}.
When there are two temporally close flare events that
have overlapping magnetograms with opposite labels, 
we give priority to the positive class, labeling the overlapping magnetograms as positive.
Table \ref{table:dataset_summary} presents the numbers of positive and negative mangetograms
obtained by our labeling scheme
for each cumulative flare class.
Note that although there are fewer positive/eruptive $\geq$C-class flares than negative/confined $\geq$C-class flares, there are more positive magnetograms than negative magnetograms 
for the $\geq$C-class flares 
as shown in Table \ref{table:dataset_summary}.
This happens because we give priority to the positive class in labeling overlapping magnetograms as indicated above, and there are many such magnetograms for the $\geq$C-class flares.
 
For each cumulative flare class,
we employed a stratified 80:20 split for training and testing our model, allocating 80\% of the magnetograms in the respective dataset shown in Table \ref{table:dataset_summary} 
to the training set and the remaining 20\% to the test set.
Magnetograms of the same AR are placed in the training set or in the test set,
but not in both. 
In this way, we ensure that our HNN model is trained with ARs
different from the ARs in the test set and makes predictions on the test ARs that the model has never seen during training.

\begin{table}
\centering
\begin{tabular}{ccc}
\hline
\hline
 Cumulative Flare Class& Positive/Eruptive    & Negative/Confined    
  \\  \hline
 $\geq$M5.0 & 2,645 & 1,685 
 \\ \hline
 $\geq$M & 6,703 & 10,631 
 \\ \hline
 $\geq$C & 12,336 & 12,059 
 \\ \hline
\end{tabular}
\caption{Sample sizes for positive and negative classifications as defined 
in our prediction task.}
\label{table:dataset_summary}
\end{table}

\subsection{The HNN Model} 
\label{subsec:Prediction Methods}

\begin{figure}
    \centering
        \includegraphics[width=0.5\linewidth]{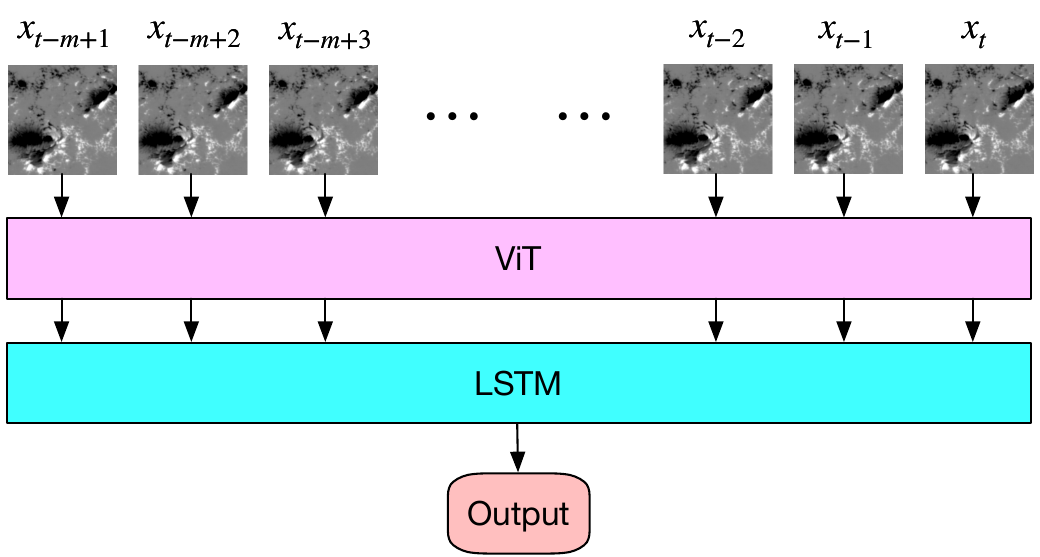} 
\caption{Overall architecture of our HNN model.
Given is a magnetogram $x_t$ at time point $t$ in an AR that will produce 
a $\gamma$-class flare within the next 24 hours of $t$.
The HNN model accepts, as input, 
a time series or
sequence of $m$ magnetograms
$x_{t-m+1}, x_{t-m+2}, x_{t-m+3} \dots, x_{t-2}, x_{t-1}, x_t$
where $m$ is set to 10.
The sequence is first processed image-by-image by a
vision transformer (ViT) 
to extract spatial features in the sequence. 
The resultant sequence of $m$ embedding vectors is then sent to a long short-term memory (LSTM) network
to capture temporal dependencies in the sequence.
Finally, the HNN model produces, as output, ``1'' indicating that 
$x_t$ is positive (i.e., the $\gamma$-class flare is eruptive)
or
``0'' indicating that $x_t$ is negative
(i.e., the $\gamma$-class flare is confined).}
\label{fig:HNN}
\end{figure}

\begin{figure}
\centering
\includegraphics[width=0.67\linewidth]{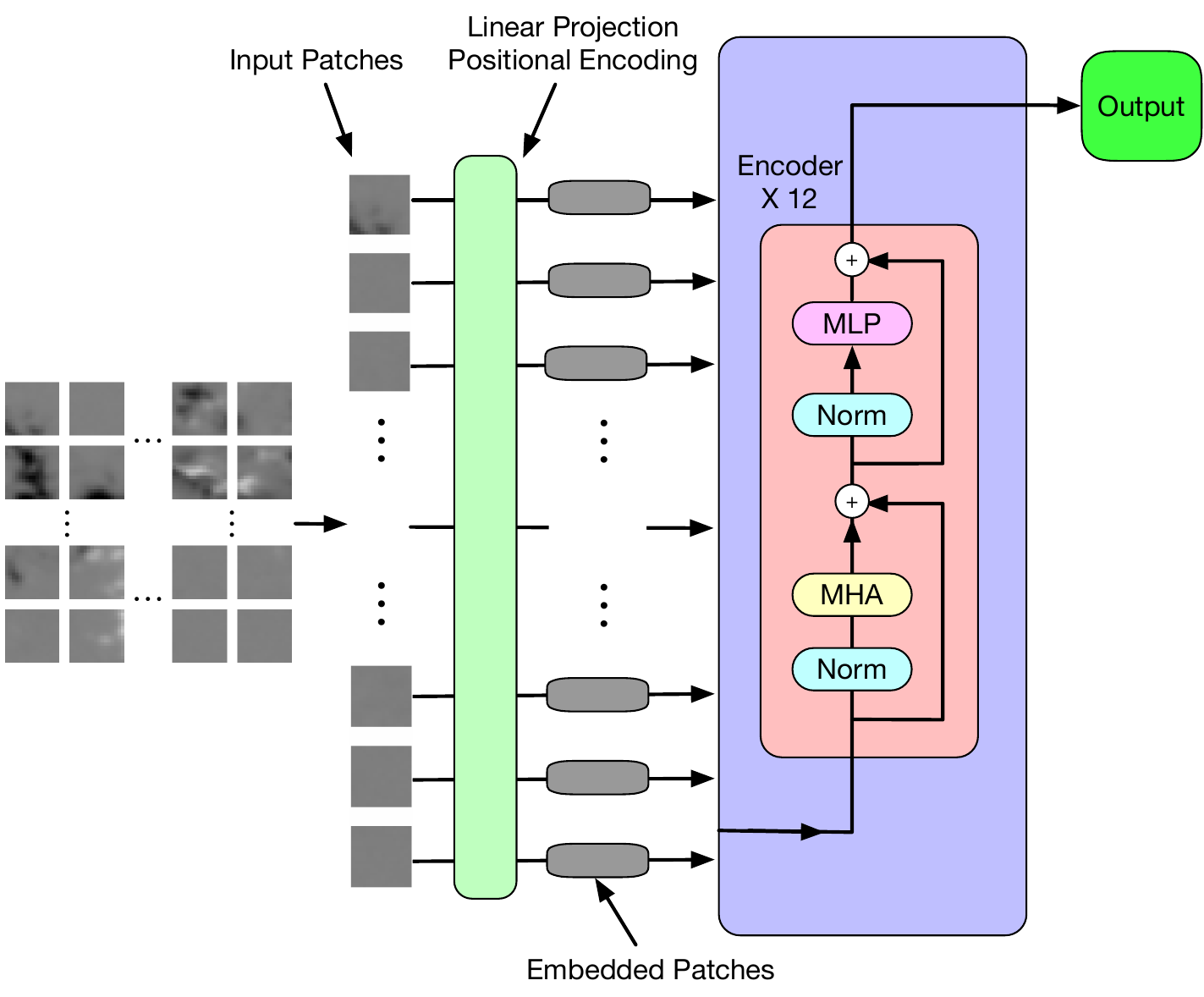} \\
\caption{Architectural design of the vision transformer (ViT).
Each input magnetogram image is divided 
into non-overlapping patches, which are embedded via linear projection and positional encoding. These patch embeddings are fed into a stack of transformer encoder layers,
each of which contains a multi-head self-attention (MHA) module, layer normalization (Norm), and 
a multilayer perceptron (MLP) module, to generate the final image representation.}
\label{fig:ViT}
\end{figure}

Figure~\ref{fig:HNN} shows the overall architecture of the proposed HNN model.
The model leverages the strength of a vision transformer
(ViT) in capturing global spatial dependencies within an image and the capability of 
long short-term memory (LSTM) for sequence modeling. 
Let $x_{t}$ represent the magnetogram collected at time point $t$ in an AR.
The input of HNN is a sequence of
$m$ consecutive magnetograms
$x_{t-m+1}, x_{t-m+2}, x_{t-m+3} \dots, x_{t-2}, x_{t-1}, x_t$
where time point $t$ and time point $t-i$ differ by $i$ hours.
(In our study, $m$ is set to 10.)
The label of these $m$ consecutive magnetograms is defined
as the label of the last magnetogram $x_{t}$.
Thus, if $x_{t}$ belongs to the positive class, then the input sequence
$x_{t-m+1}, x_{t-m+2}, x_{t-m+3} \dots, x_{t-2}, x_{t-1}, x_t$
is defined as positive; otherwise, the sequence is defined as negative.
Note that we collect and label 24
magnetograms
that are within the 24 hours prior to the peak time
of a flare, as illustrated in Figure \ref{fig:label_rule}.
These collected labeled magnetograms 
(positive vs. negative)
constitute the 
datasets at hand as shown in Table 
\ref{table:dataset_summary}.
When we 
train/test our HNN model,
we take each magnetogram $x_t$ in 
the corresponding training/test set,
use $x_{t}$ and its preceding ($m-1$) = 9 magnetograms to
form a time series or
sequence of 10 magnetograms, and input the sequence to
the HNN model as shown in Figure \ref{fig:HNN}.
Forming this sequence of 10 magnetograms allows our model
to understand the evolution of magnetic fields and
capture temporal patterns in the
input sequence, hence enhancing model accuracy.
When there are more than two consecutive magnetograms missing in the sequence, 
$x_{t}$ is excluded from the study.
When there is only one 
or two consecutive magnetograms missing in the sequence,
we use linear interpolation to generate synthetic magnetograms, fill in the gaps, and produce a complete, non-gapped time series used as the input of HNN.

As illustrated in Figure~\ref{fig:HNN}, 
the $m$ input magnetograms are sent to the 
vision transformer (ViT),
which captures the spatial patterns in the input magnetograms and outputs $m$ embedding vectors, one embedding vector corresponding to one input magnetogram.
The sequence of $m$ embedding vectors is then sent to 
the LSTM network,
which captures the temporal patterns in the embedding vector sequence.
Finally, our HNN model outputs ``1'' indicating that $x_{t}$ is positive
(that is, a flare-associated CME will be triggered)
or ``0'' indicating that $x_{t}$ is negative
(that is, a flare-associated CME will not be triggered).

 Figure~\ref{fig:ViT} presents 
 the architectural design
 of the vision transformer (ViT).
 The input of the ViT is a magnetogram, which is divided into image patches of $N \times N$ pixels.
(In the study presented here, $N$ is set to 16.)
Each image patch is flattened into a vector, 
which is linearly projected onto an embedding space
by a linear projection layer.
To retain spatial information, which is lost during the patching process, 
positional embeddings are added to the patch embeddings
by a positional encoding layer.
These positional embeddings are typically learnable parameters that allow the ViT to understand the original location of each patch. 
The resultant sequence of embedding vectors is then sent to the ViT encoders.

The ViT contains $E$ encoders.
(In our study presented here, $E$ is set to 12.)
Each encoder updates the sequence representation based on two modules:
a multi-head self-attention (MHA) module
where the number of attention heads in an encoder is
also set to 12, 
and a position-wise multilayer perceptron (MLP) module.
Layer normalization \citep{LeCun2015} is applied before each module and a residual connection \citep{DBLP:conf/cvpr/HeZRS16},
denoted by $\oplus$, is used after each module.
The MHA mechanism allows the ViT to dynamically weigh the relevance of different patches, capturing long-range dependencies
throughout the input magnetogram by computing the attention scores between all pairs of patch embeddings. 
The MLP module consists of two linear layers 
with a Gelu non-linearity in between, applied independently to each sequence element. This deep processing
allows the ViT to extract complex hierarchical features from the spatial arrangement of patches.
To accelerate convergence and improve model performance,
we load the pre-trained ViT 
weights of \citet{DBLP:conf/iclr/DosovitskiyB0WZ21}
to our ViT model.
This allows the model to leverage the features
learned from large-scale datasets, thereby improving generalization and training efficiency.

After processing the data through the $E$ encoders, 
we produce a final embedding vector, which is the output of the ViT.
There are $m$ input magnetograms, 
which are sent to the ViT.
The ViT outputs a sequence of $m$ embedding vectors.
The embedding vector sequence is then fed to
the LSTM network as shown
in Figure~\ref{fig:HNN}. 
The LSTM network processes this sequence to further refine features and capture potential sequential patterns within the embedded spatial data, leveraging its gating mechanisms to model complex dependencies.

The final output of the LSTM network in
Figure~\ref{fig:HNN} 
is sent to a classification layer,
which predicts the probability $\hat{y}$ of the occurrence of a flare-associated CME.
Our primary goal is binary classification (that is,
there will be an eruptive flare vs.
there will be a confined flare within the next 24 hours).
When $\hat{y}$ is greater than or equal to the default
probability 0.5, the output is ``1'' or positive,
indicating that there will be an eruptive flare
within the next 24 hours.
When $\hat{y}$ is smaller than 0.5, the output is ``0'' or negative,
indicating that there will be a confined flare
within the next 24 hours.
Recognizing the
class imbalance inherent in our datasets, as shown in Table \ref{table:dataset_summary},  
we adopt a class-balanced focal loss \citep{DBLP:conf/iccv/LinGGHD17}, 
which is defined as:
\begin{equation}
\label{eq:focal_loss}
\text{FL}(y, \hat{y}) = -\alpha_s (1-p_s)^\lambda \log(p_s),
\end{equation}
where $y \in \{0, 1\}$ is the ground truth label, 
$\hat{y}$ is the model's estimated probability for class $y=1$, and
\begin{eqnarray}
    p_s &= \begin{cases} \hat{y} & \text{if } y=1 \\ 1-\hat{y} & \text{if } y=0 \end{cases} 
    \end{eqnarray}
   \begin{eqnarray} 
    \alpha_s &= \begin{cases} \alpha & \text{if } y=1 \\ 1-\alpha & \text{if } y=0 \end{cases}
\end{eqnarray}
Here, $\alpha \in [0, 1]$ is a balancing factor to address class imbalance (weighting the minority class higher), and 
$\lambda \ge 0$
is the focusing parameter that down-weights well-classified examples, forcing the model to focus on harder, misclassified examples. 
(In the study presented here, $\alpha$ is set to 0.4 and $\lambda$ is set to 2.)
The focal loss function helps mitigate the dominance of the majority class during training.
The parameters of the HNN model are optimized using the Adam optimizer
\citep{DBLP:journals/corr/KingmaB14}
with a learning rate of 0.0001.
The batch size is 32 and the number of epochs is 40.

\section{Results} \label{sec:Results}

\subsection{Experimental Setup and Evaluation Metrics}
Given an AR that produces a $\gamma$-class flare within the next 24 hours of a time point $t$ and a magnetogram
$x_{t}$ collected at time point $t$,
we define $x_{t}$ as a true positive (TP) if 
our HNN model predicts that $x_{t}$ is positive, 
and $x_{t}$ is indeed positive, i.e., the $\gamma$-class flare is associated with a CME.
We define $x_{t}$ as a false positive (FP) if
our model predicts that $x_{t}$ is positive
while $x_{t}$ is actually negative, i.e., the $\gamma$-class flare is not associated with a CME. 
We say $x_{t}$ is a true negative (TN) if 
our model predicts $x_{t}$ to be negative
and $x_{t}$ is indeed negative;
$x_{t}$ is a false negative (FN) if 
our model predicts $x_{t}$ to be negative
while $x_{t}$ is actually positive.
We also use TP (FP, TN, FN, respectively) to represent the total number of true positives (false positives, true negatives, false negatives, respectively)
produced by our model.

The evaluation metrics used in the study include the following:
\begin{align}
\text{Recall} & = \frac{\mathrm{TP}}{\mathrm{TP} + \mathrm{FN}}, 
\\
\text{Precision} & = \frac{\mathrm{TP}}{\mathrm{TP} + \mathrm{FP}}, \\
\text{Accuracy (ACC)} & = \frac{\mathrm{TP} + \mathrm{TN}}{\mathrm{TP} + \mathrm{FP} + \mathrm{TN} + \mathrm{FN}}, \\
\text{Heidke Skill Score (HSS)} & = \frac{2(\mbox{TP}\times \mbox{TN}-\mbox{FP}\times \mbox{FN})}{(\mbox{TP+FN})(\mbox{FN+TN})+(\mbox{TP+FP})(\mbox{FP+TN})},\\
\text{True Skill Statistics (TSS)} & = \frac{\mathrm{TP}}{\mathrm{TP}+\mathrm{FN}} - \frac{\mathrm{FP}}{\mathrm{TN}+\mathrm{FP}}.
\label{eqn:TSS}
\end{align}

In the above, recall, precision and accuracy range between  
0 and 1,
with higher values indicating better performance. HSS and TSS range between 
$-1$ and 1.
HSS shows the performance in comparison to a generic random prediction (negative/positive values meaning worse/better performance), while TSS compares the true positive rate (first term in 
Equation (\ref{eqn:TSS}))
with the false positive rate
(second term in Equation (\ref{eqn:TSS})). 
HSS = 0 implies performance identical to that of a 
no-skill, generic random prediction, while TSS = 0 indicates an exact balance between the true positive rate and the false positive rate. 
For a comprehensive account, see 
\citet{JS-2011} and
\citet{2012ApJ...747L..41B}.

We use the vision transformer (ViT) alone as a baseline method.
Unlike our HNN model, which uses a time series or sequence of
$m$ consecutive magnetograms
$x_{t-m+1}, x_{t-m+2}, x_{t-m+3} \dots, x_{t-2}, x_{t-1}, x_t$
as input to predict whether $x_{t}$ is positive or negative, the ViT baseline method only uses $x_{t}$ as input
to make the prediction.
Thus, the ViT baseline method captures the spatial patterns
of the input magnetogram $x_{t}$ without considering the temporal patterns
in the magnetogram sequence used by our HNN model.

\subsection{Performance Evaluation of the HNN Model}
\begin{figure}
\begin{center}
\includegraphics[height=0.43\textheight]{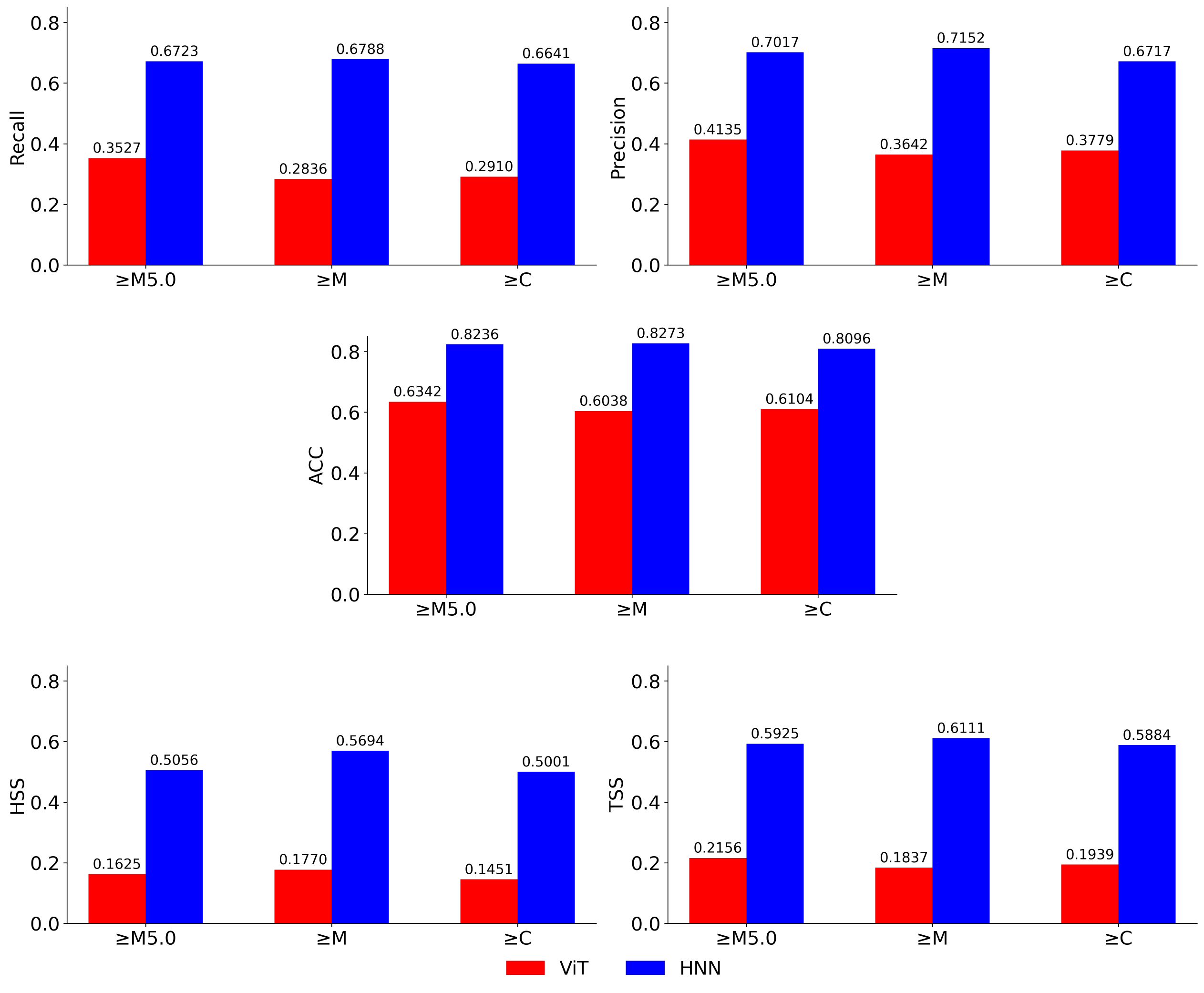}
\caption{Performance comparison 
between ViT and HNN.}
\label{fig:bar_chart}
\end{center}
\end{figure}

Figure \ref{fig:bar_chart} 
evaluates the performance of HNN and compares it with ViT.
It can be seen in Figure \ref{fig:bar_chart} that
HNN outperforms ViT in all five metrics.
Specifically, for the
$\geq$M5.0-class flares,
HMM achieves a TSS of 0.5925 compared to the TSS of 
0.2156 obtained by ViT.
For the $\geq$M-class flares,
HMM achieves a TSS of 0.6111 compared to the TSS of
0.1837 obtained by ViT.
For the $\geq$C-class flares,
HNN achieves a TSS of 0.5884 compared to the TSS of
0.1939 obtained by ViT.
These findings underscore the effectiveness of combining ViT with LSTM in 
our HNN framework. Good performance in recall, precision, accuracy, HSS and TSS with solar observations as direct input without  
handpicked
features highlights the potential of HNN as an effective tool for operational space weather forecasting.

\begin{figure}
    \centering
    \begin{minipage}{0.65\textwidth}
        \centering
        \includegraphics[width=\linewidth]{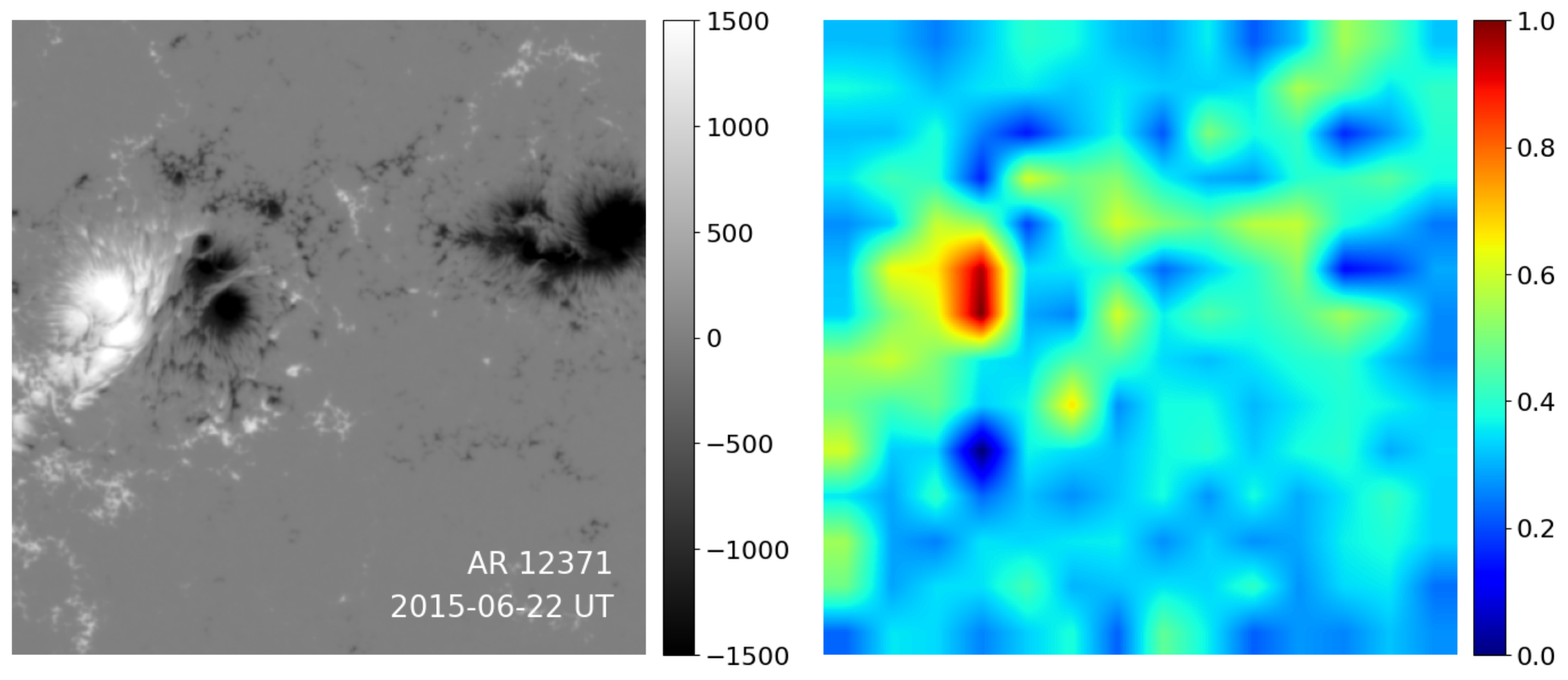} \\
        \textbf{(a)}
         \vspace*{+0.5cm}
    \end{minipage}
    \begin{minipage}{0.65\textwidth}
        \centering
        \includegraphics[width=1\linewidth]
        {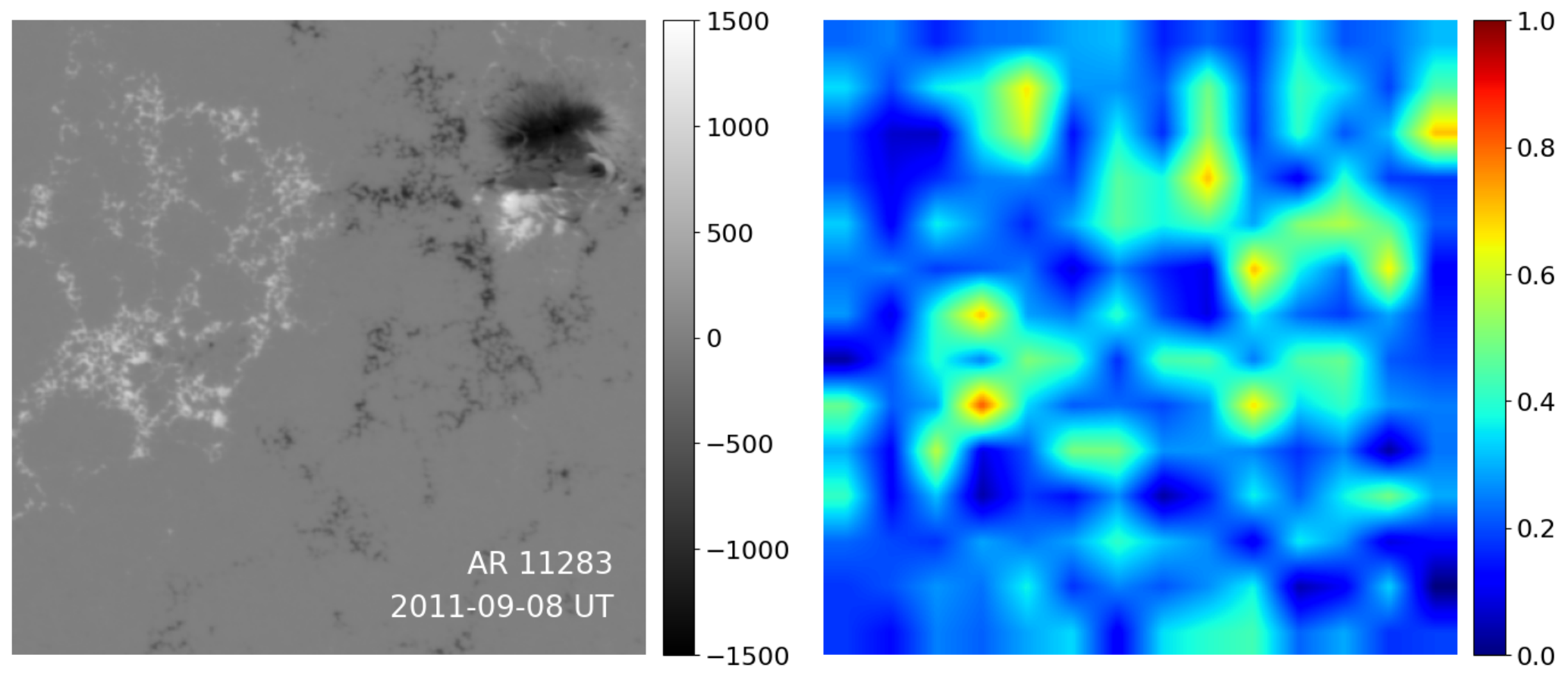} \\
        \textbf{(b)}
                 \vspace*{+0.3cm}
    \end{minipage}
    \caption{Attention heat maps of the HNN model for two test magnetograms.
     The grayscale bar corresponds to the LOS magnetic field, in Gauss.
     The color bar shows the attention score.
    A larger attention score at a region indicates that more attention is paid to the region, where large attention scores are represented by dark red and
small attention scores are represented by dark blue.
(a) A positive prediction 
(true positive) 
where the magnetogram on the left is predicted to be positive, related to an eruptive flare. The model pays much attention to the 
vicinity
of a polarity inversion line (PIL) region in the magnetogram.
(b) A negative prediction 
(true negative) 
where 
the magnetogram on the left is predicted to be negative, related to a confined flare.
The model pays no attention to 
the clearly weaker PIL region in the magnetogram.
Instead, the model pays some attention to the region where no PIL exists.}
\label{fig:maps}
\end{figure}

\begin{figure}
\centering
\includegraphics[width=0.67\linewidth]{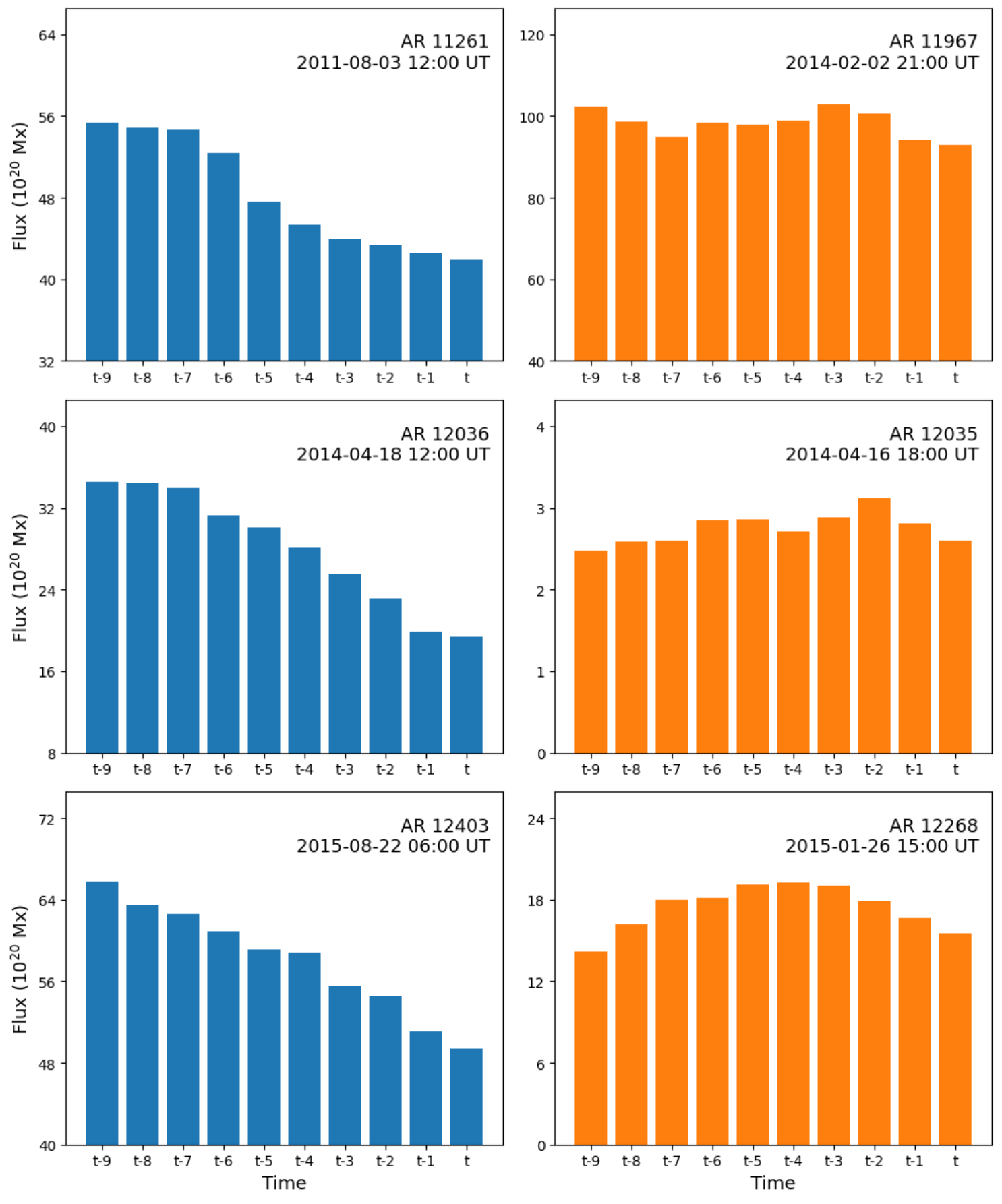} \\
\caption{
Illustration of the changes of the unsigned magnetic flux in the PIL regions
of the 10 magnetograms at time points
$t-9, t-8, \ldots, t-1, t$
used as the input of our HNN model
in six ARs
where $t$ is as specified in each panel.
Left panels: results for three positive predictions 
(true positives), where a positive
prediction indicates that the corresponding AR will produce an 
eruptive flare within the next 24 hours of $t$.
Right panels: results for three negative predictions
(true negatives), 
where a negative
prediction indicates that the corresponding AR 
will produce a 
confined flare within the next 24 hours of $t$.
In the three positive predictions,
the unsigned magnetic flux in the PIL regions decreases in the course of time, which may point to magnetic flux cancellation.}
\label{fig:fluxcancel}
\end{figure}

\begin{figure}
\centering
\includegraphics[width=0.67\linewidth]{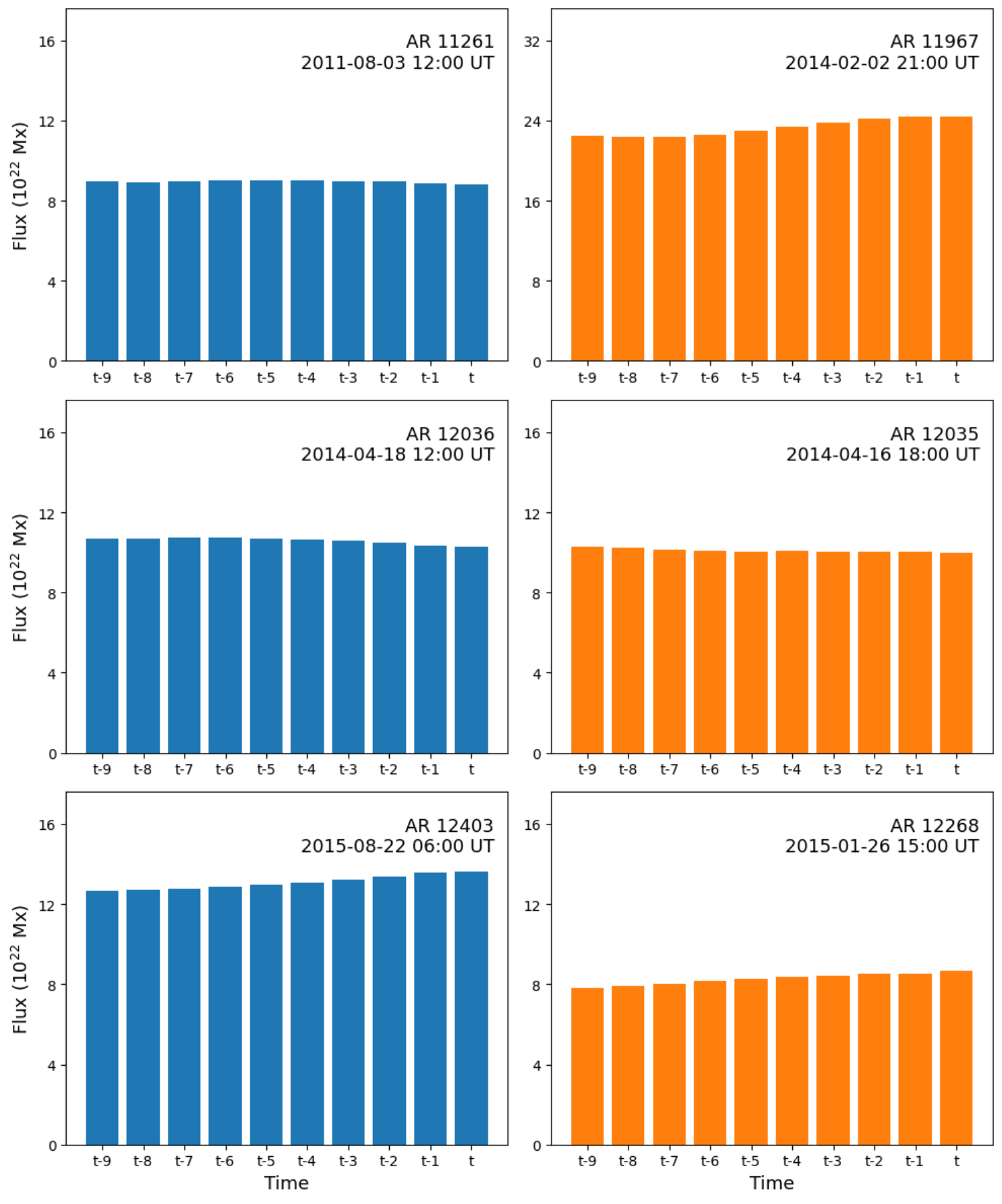} \\
\caption{Illustration of the changes of the total unsigned magnetic flux in the
10 magnetograms at time points
$t-9, t-8, \ldots, t-1, t$ 
used as the input of our HNN model
in the six ARs
shown in Figure \ref{fig:fluxcancel}
where $t$ is as specified in each panel.
Left panels: results for the three positive predictions 
(true positives), where a positive
prediction indicates that 
the corresponding AR will produce an 
eruptive flare within the next 24 hours of $t$.
Right panels: results for the three negative predictions
(true negatives), 
where a negative
prediction indicates that 
the corresponding AR 
will produce a 
confined flare within the next 24 hours of $t$.
There is no clear difference in the trends between the results for the three positive predictions and those for the three negative predictions.}
\label{fig:CMF}
\end{figure}

\subsection{Model Interpretation}
HNN is a transformer-based model with attention mechanisms.
To elucidate the decision-making process of our model, 
we employ saliency methods 
\citep{DBLP:conf/bmvc/DjilaliMO23,DBLP:journals/jaihc/XuYDL23}
that use the attention weights within the transformer layers 
of the model to infer importance
and directly link the model's predictions to the most influential
regions in the input magnetogram. 
We leverage the attention mechanisms and generate
attention heat maps, which reveal the regions where the model focuses the most during inference.
Figure \ref{fig:maps}(a) (Figure \ref{fig:maps}(b), respectively) presents an attention heat map
for a positive (negative, respectively) prediction.
In Figure \ref{fig:maps}(a), 
according to the attention heat map in the right panel, 
our model pays much attention to the 
vicinity of a polarity inversion line (PIL) region in the magnetogram
in the left panel.
As a result, the model predicts that
the magnetogram is positive, 
i.e., a flare will occur within the next 24 hours and the flare is eruptive.
This prediction result is consistent with
the established fact that CMEs
 originate near or along PILs
\citep{2019A&A...626A..91L,2024MNRAS.534..444A}.
In Figure \ref{fig:maps}(b), 
according to the attention heat map in the right panel, 
our model does not pay any attention to the PIL region in the 
magnetogram in the left panel. 
Instead, the model pays some attention to a region
where no PIL exists.
As a consequence, the model predicts that the 
magnetogram is negative, i.e., a flare will occur within the next 24 hours 
and the flare is confined.
These attention heat maps help us better understand the decision-making process of
the HNN model.
The results offer enhanced reliability of HNN for distinguishing between eruptive and confined flare events.

\subsection{Physics-based Analysis of Model Predictions}

We further
pursue a physics-based analysis of model predictions, 
examining the relationship between
magnetic flux cancellation and the occurrence of 
flare-associated CMEs.
Magnetic flux cancellation 
occurs along PILs where 
magnetic fields 
of opposite polarity 
frequently come into contact.
We consider six ARs where each AR is
associated 
with a reference time point $t$ (Figure \ref{fig:fluxcancel}).
Our HNN model predicts whether the magnetogam at $t$ 
(i.e., $x_{t}$) in the AR
is positive 
(i.e., a flare will occur within the next 24 hours of $t$ and the flare is eruptive)
or negative
(i.e., a flare will occur within the next 24 hours of $t$ and the flare is confined).
As described in Section \ref{subsec:Prediction Methods},
the input of the HNN model is a sequence of
10 consecutive magnetograms
$x_{t-9}, x_{t-8}, \dots, x_{t-1}, x_{t}$.
Magnetic flux cancellation can be quantified by 
the unsigned magnetic flux contained within the PIL regions
in each magnetogram.
Using the PIL identification algorithm developed by \citet{2023ApJS..265...28J},
we locate all PIL regions in a magnetogram.

Figure \ref{fig:fluxcancel} shows the changes of 
the unsigned magnetic flux in the PIL regions 
of the 10 magnetograms 
used as the input of our HNN model
in each of the six ARs. 
The left panels with
blue histograms
show three positive predictions (true positives)
where a positive
prediction indicates that the corresponding AR will produce an eruptive flare.
The right panels with
orange histograms
show three negative predictions (true negatives)
where a negative
prediction indicates that the corresponding AR will
produce a confined flare.
It is evident that, in the three positive predictions, the unsigned magnetic flux in the PIL regions decreases in the course of time, which may point to magnetic flux cancellation. We must emphasize, of course, that since we only rely on the LOS field component, similar to the vertical magnetic field if the target region is relatively close to the central meridian, such a behavior could also be caused by a systematic change in the geometry of the photospheric magnetic field vector, i.e., becoming more horizontal in the course of time. Fully establishing that magnetic flux cancellation plays a role in the behavior seen in Figure 
\ref{fig:fluxcancel}
(blue histograms) implies a detailed analysis of the full magnetic field vector and EUV/X-ray transient brightenings along the PIL regions, which exceeds the scope of this work. We note, however, that flux cancellation may well play a role, as the behavior reported by 
\citet{2023ApJ...942...27L} 
points to a systematic weakening of the photospheric horizontal field in the hours prior to eruptive flares (but not prior to confined flares), only to return to initial pre-eruption level values immediately after the eruptions. Showing here that the vertical magnetic field becomes weaker instead of strengthening that would be expected by the results of 
\citet{2023ApJ...942...27L}
indicates true loss of flux, possibly due to flux cancellation. 
However, for the three negative predicitons (orange histograms),
no clear trend is observed in the changes of the unsigned magnetic flux in the PIL regions.

We next examine the relationship between the 
larger-scale magnetic environment in an AR and the occurrence of 
flare-associated CMEs.
Specifically, 
in each magnetogram, 
$x_{t-9}, x_{t-8}, \dots, x_{t-1}, x_{t}$,
of the AR,
we compute the total unsigned magnetic flux of the whole magnetogram.
Figure \ref{fig:CMF}
shows the changes of 
the total unsigned magnetic flux in the 10 magnetograms
used as the input of our HNN model
in each of the six ARs 
in Figure \ref{fig:fluxcancel}. 
There is no clear difference in the trends between the resuls for the three positive predictions 
(blue histograms)
in the left panels and those for the three negative predictions
(orange histograms)
in the right panels in Figure \ref{fig:CMF}.

\section{Discussion and Conclusion}
\label{sec:conclusion}

We developed a hybrid neural network (HNN) to 
predict whether a $\gamma$-class flare that will occur within the next 24 hours will be 
associated with
a CME, i.e., whether the flare will be eruptive or confined.
We considered $\gamma$ 
to be defined cumulatively, as
$\geq$M5.0, $\geq$M,
or $\geq$C.
When $\gamma$ was $\geq$M, the HNN model achieved a TSS of
0.6111.
A similar result with slightly lower TSS value, albeit using different data sets and prediction methods, was reached by \citet{2020ApJ...890...12L},
though \citet{2020ApJ...890...12L} did not
consider $\geq$M5.0/$\geq$C-class flares.
Note that the TSS value is calculated based on the test set at hand.
If the methodology presented here was to be used in a near-real time operation scheme, 
its testing results on eruptive flares versus confined flares may be different than the ones presented here. 
Note also that, as in the work of \citet{2020ApJ...890...12L},
we rely on existing tools
\citep[e.g.,][]{2015ApJ...798..135B,
2017ApJ...843..104L,
2018SoPh..293...28F,
2018ApJ...856....7H,
2018SoPh..293...48J,
2018ApJ...858..113N,
2019ApJ...877..121L,
2020ApJ...891...10L,
2020ApJ...895....3W,
2022ApJ...941....1S,
2022ApJS..263...28Z,
2023NatSR..1313665A,
2025ApJS..277...60W,
2025ApJS..276...68X}
to predict whether a $\gamma$-class flare
will occur within the next 24 hours.
If it is predicted that a $\gamma$-class flare will occur,
then one can use the HNN model developed here to further predict
whether that flare will be eruptive (i.e., it 
is associated with
a CME) or confined
(i.e., it is not associated with a CME).
It should be pointed out that our work
is based on DONKI.
There are other data sets for flares and CMEs
\citep[e.g.,][]{2025DIB....6011539J}.
It is worthwhile to consider those other data sets in future studies.

Our HNN model employs several parameters and hyperparameters.
For example, we use a time series or sequence of 10 magnetograms as input to the HNN model.
We tried shorter sequences with 6 or 8 magnetograms 
and longer sequences with
14 or 18 magnetograms as input.
Shorter sequences tended to underutilize the available temporal information in the input, whereas longer sequences increased computational time without yielding better performance. 
The input magnetogram to the ViT of our model is
divided into image patches of $16 \times 16$ pixels.
We also tried smaller patches of $8 \times 8$ pixels and
larger patches of $32 \times 32$ pixels.
Smaller patches yielded more patch embeddings and hence increased processing time while performance is not improved.
Larger patches caused a loss of
fine-scale magnetic structure information,
yielding worse performance.
Likewise, the ViT uses 12 encoders.
ViT modules with 
 fewer encoders tended to underfit the training data, while ViT modules with more encoders and deeper configurations increased training time without yielding better performance.
These hyperparameter values are optimized
using the grid search strategy 
\citep{10.5555/1953048.2078195}.

To explain a prediction made by our HNN model, we
leveraged the attention mechanisms of the model
using saliency methods and generated
attention heat maps, which revealed the regions
in an input mangetogram where the model focused the most during inference.
For a positive prediction (that is, a prediction
indicating that the flare related to the input mangetogram was eruptive),
the heat map indicated that the model
paid much attention to the vicinity of
polarity inversion lines (PILs) in the input magnetogram.
For a negative prediction (that is,
a prediction indicating that the flare related to the input mangetogram was confined), 
the heat map 
indicated that the model paid attention to
regions where no PIL existed.
The heat maps provided insight into
how our model produced its predictions, making the model more interpretable.
In addition, 
we aimed to physically understand our model predictions. 
It appears likely
that magnetic flux cancellation
in PIL regions 
plays a role
in triggering flare-associated CMEs,
a finding consistent with literature
\citep{2018ApJ...864...68S,
2023ApJ...942...27L,
2025A&A...699A..87P}.
This said, the evolution of the total unsigned magnetic flux in the entire active region does not appear to show a clear relation to eruptive flares.

\begin{figure}
\begin{center}
\includegraphics[height=0.43\textheight]{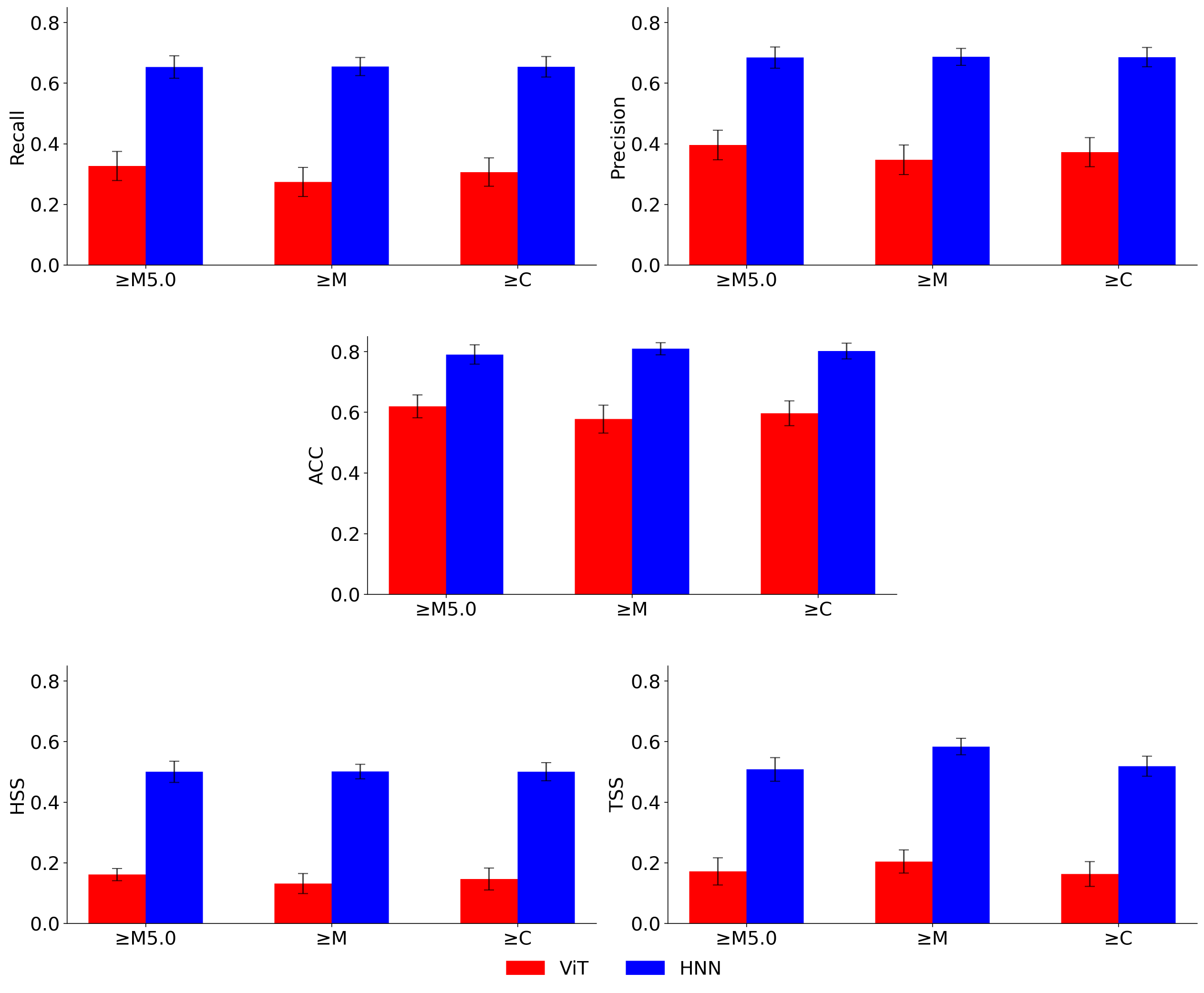}
\caption{Bar graphs showing the comparison between ViT and HNN based on the five-fold cross-validation scheme.}
\label{fig:bar_chart_5f}
\end{center}
\end{figure}

To further understand the behavior of our HNN model and
compare it with the baseline method (ViT),
we conducted an additional five-fold cross-validation experiment.
In this experiment,
there were five runs.
In each run, 
for each cumulative class of flares, 20\% of the data in
Table \ref{table:dataset_summary} were selected from the respective dataset and used
for testing; the remaining 80\% of the data were used for training.
Again, magnetograms of the same AR were placed in the training set or in the test set,
but not in both. 
We recorded recall, precision, accuracy (ACC),
Heidke Skill Score (HSS),
and True Skill Statistic (TSS) in each run and
calculated the mean and standard deviation
over the five runs. 
Figure \ref{fig:bar_chart_5f} 
presents the results,
where each
colored bar represents the mean of the five runs and its
associated error bar represents the standard deviation divided by
the square root of the number of runs.
The figure shows that
HNN consistently outperforms ViT
in all metrics,
consolidating the findings in Figure \ref{fig:bar_chart}.
On the basis of these results, we conclude that the HNN model  
shows promise
for predicting associations between
solar flares and CMEs. 

\begin{acknowledgments}
We acknowledge constructive suggestions and comments from an anonymous referee and the scientific editor that have helped to significantly improve the manuscript.
We also thank members of the Institute for Space Weather Sciences
for fruitful discussions.
SDO is the first mission launched for NASA's Living With a Star (LWS) Program. The SDO/HMI data is provided by the Joint Science Operations Center (JSOC) 
Science Data Processing (SDP).
DONKI is developed by the Community Coordinated Modeling Center (CCMC) at NASA. 
The proposed HNN model is 
implemented in PyTorch and Astropy packages.
V.Y. acknowledges support from
NSF AGS grants
2401229,
2408174,
2300341,
2309939,
NSF AST-2108235 grant, and
NASA 80NSSC24K1914 grant.
J.W. and H.W. acknowledge support from 
NSF AGS grants
2228996, 
2149748, 
NSF OAC grants
2504860,
2320147, 
and NASA grants 
80NSSC24K0548,
80NSSC24K0843, 
80NSSC24M0174.
M.K.G. acknowledges support from NSF OAC grants 2504861 and 2513887.
Y.X. acknowledges support from NSF AGS grants
2228996,
2229064, 
and NSF 
RISE-2425602 grant.
\end{acknowledgments}

\bibliographystyle{aasjournal}

\end{document}